\documentstyle[prd,preprint,aps,epsf]{revtex}

\newcommand{\xvec}{{\bf x}}

\newlength{\headroom}
\newlength{\psfigskip}
\def\xvec{{\bf x}}
\def\kvec{{\bf k}}
\def\rvec{{\bf r}}

\begin{document}

\title{
		     Cosmic string formation from correlated fields
}
\author{             Robert\ J.\ Scherrer, }
\address{
{\it
		     Department of Physics and Department of Astronomy,
		     The Ohio State University,               \\
	             Columbus, Ohio 43210                     \\
}}
\author{             Alexander Vilenkin        }
\address{
{\it
                     Institute of Cosmology,
                     Department of Physics and Astronomy,     \\
                     Tufts University, Medford, MA 02155        \\
}}

\maketitle
%

\renewcommand{\baselinestretch}{1.3}
\begin{abstract}

We simulate the formation of cosmic strings at the zeros of
a complex Gaussian field with a power spectrum $P(k) \propto k^n$,
specifically addressing the issue of the fraction of length in
infinite strings.  We make two improvements over previous
simulations:  we include a non-zero random background field in our box
to simulate the effect of long-wavelength modes, and we examine
the effects of smoothing the field on small scales.  The
inclusion of the background field
significantly reduces the fraction of length in infinite
strings for $n < -2$. Our results are consistent with
the possibility that
infinite strings disappear at some $n =  n_c$ in the range
$-3 \le n_c < -2.2$, although we cannot rule out $n_c = -3$, in which
case infinite strings would disappear only at the point where the
mean string density goes to zero.
We present an analytic argument which suggests the latter case.
Smoothing
on small scales eliminates closed loops on the order of the lattice
cell size and
leads to a ``lattice-free" estimate of the infinite
string fraction.  As expected, this fraction 
depends on the type of window function used for smoothing.

\end{abstract}
\pacs{PACS numbers: }
%


\section{Introduction}

Cosmic strings are effectively one-dimensional topological defects
which may form at a phase transition in the early universe
(see Ref. \cite{review} for a review).  Although much
of the early interest in cosmic strings has centered on
the possibility that they might have served as seeds for the
formation of large-scale structure, cosmic strings
are interesting physical objects in any case, and they have
analogues in the study of condensed matter \cite{condmat}.

Any investigation of the evolution and cosmological consequences
of cosmic strings must begin with the study of the initial
cosmic string configuration, a study which was first undertaken
by Vachaspati and Vilenkin \cite{VV}, and re-examined by
many others (\cite{SF} - \cite{RY}).
Although subsequent cosmic
string evolution will erase many of the details of the initial
configuration, one fundamental property of the initial conditions
is crucial to the subsequent evolution:  the existence of infinite
strings.  Without the existence of infinite strings,
the cosmic strings produced at the phase transition may all
decay via gravitational radiation long before they can have
any interesting cosmological effects.  Vachaspati and Vilenkin
found in their simulation that roughly 80\% of the string was
in the form of infinite strings \cite{VV}.
Using a very different type of simulation for the string formation
process, Borrill \cite{Borrill} claimed that this fraction was, in fact,
zero, while Robinson and Yates \cite{RY}, in a study of the dependence
of this fraction on the power spectrum of the initial field,
argued that for power spectra of the form $P(k) \propto k^n$,
the infinite string fraction $f_\infty$ drops to zero for $n \le -2$.
In fact, this result is not obvious from their simulations;
it is based on fitting $f_\infty$ to an analytic function
for $n > -2$.

In this paper, we extend the simulations of
Robinson and Yates in two ways.  First, we include the effects
of long-wavelength modes which are absent from earlier simulations.
Second, we examine the effect of smoothing the initial field
to remove lattice effects.  We find that the former
change has a dramatic effect on $f_\infty$,
sharply reducing $f_\infty$ for $n < -2$.  Smoothing
also affects $f_\infty$, leading to ``lattice-free" estimates
of the fraction in infinite strings.  In the next section we
present our numerical results, and in Section 3 we discuss
briefly our main conclusions.

\section{Numerical simulations}

With one exception \cite{Borrill}, simulations of cosmic string
formation typically make use of a lattice.  A value
of the field $\phi$ associated with
the string is assigned to each of the
cells of the lattice, and the location of the
cosmic string is then identified with edges of the lattice
around which the field winds through 360$^o$.  The original
simulations of Vachaspati and Vilenkin \cite{VV} were performed
on a cubic lattice with no correlations between values of the field
in different cells.  Subsequent researchers investigated changing
the probability of string formation by ``biasing" the distribution
of the values of the field
\cite{vachaspati}, \cite{kibble}, \cite{HS},
by allowing the
field to be divided into domains of variable size \cite{kibble}, or
by allowing long-range correlations between the field
values \cite{RY}.  All simulations on a cubic lattice
suffer from the problem that an ambiguity exists with
regard to the string assignments at the vertices of the cubes;
this problem can be eliminated by going to more exotic
lattices, such as the tetrakaidekahedral lattice, for
which only four edges meet at every vertex \cite{SF},\cite{HS}.

In this paper, we simulate the formation of cosmic strings
using a cubic lattice and a complex Gaussian field $\phi$ with long-range correlations,
where the strings are taken to lie along the zeros of $\phi$.
A Gaussian field is completely characterized
by its power spectrum, defined by
\begin{equation}
P(k) = \int d^3 r~ e^{i \kvec\cdot \rvec} \langle \phi(\xvec)
\phi(\xvec+\rvec)\rangle.
\end{equation}
We take $\phi$ to have a power-law power spectrum:
\begin{equation}
\label{power}
P(k) \propto k^n.
\end{equation}
Models of this sort were first examined by Vishniac, Olive,
and Seckel \cite{VOS}, who proposed that cosmic strings with
$n=-3$ could be produced due to quantum fluctuations of the field
$\phi$ during inflation.
Vishniac et al. showed that the
mean string length per unit volume, $L/V$ is given by
\cite{VOS}
\begin{equation}
\label{L/V}
L/V = {1 \over 3 \pi} \langle k^2 \rangle,
\end{equation}
where
\begin{equation}
\label{k2}
\langle k^2 \rangle = {\int P(k) k^4 dk \over \int P(k) k^2 dk}.
\end{equation}
Note that some sort of smoothing or cutoff is required
for convergence
at large $k$, but this is provided automatically by the lattice
cut-off in numerical simulations.  As $n \rightarrow -3$,
equation (\ref{L/V}) gives $L/V \rightarrow 0$; this result
is confirmed by our numerical simulations.  Physically, what happens
is that within arbitrarily large regions of space
the field $\phi$ has Re($\phi) > 0$ (for example).
An expression similar to equation (\ref{L/V})
is also given in reference \cite{RY}.

In our model,
we assign values of $\phi$ to the sites on a cubic periodic lattice,
and the
string is assumed to lie along the location of the zeros of this complex field.
This model,
as we have described it, is identical to that of reference \cite{RY},
but we make two improvements in the model, both of which
have to do with the limitations in dynamic range inherent in such
a simulation.

Consider first the largest scales.  Because our simulation volume
is finite, the
simulation loses power on all scales larger than the box size,
a problem which was noted by Robinson and Yates \cite{RY}.
In the simulations, this corresponds to the fact that the mean
value of $\phi$ averaged over the entire box is zero.
This problem can be resolved by adding a random uniform background field
$\phi_b$ to every cell in the box; this field represents the contribution
to the field value from all Fourier modes which are larger than
the box size.  The variance of $\phi_b$ can be determined from
equation (\ref{power}), and in our simulations it is chosen to have
a Gaussian distribution with this variance.

Our second addition to the simulation involves the behavior of the field
on small scales.  The use of a lattice to simulate string
formation is obviously unphysical, and numerical simulations clearly
show that $f_\infty$ depends on the lattice being used.
(Compare, e.g., the results of reference \cite{VV} for a cubic lattice
with those of \cite{SF} and \cite{HS} for a tetrakaidekahedral lattice.)
On the other hand, we do expect
the field $\phi$ to be correlated on the smallest scales, leading
to some sort of domain structure.   We resolve this problem by
smoothing the field on a scale larger than the cell size to eliminate
lattice effects.

Our simulations were performed on a cubic lattice of size $128^3$.
A Gaussian, complex-valued random field $\phi$ having the power-law
power spectrum
given in equation (\ref{power})
was set down on the lattice.
The strings were identified as vortices of the field $\phi$.
In tracing a closed path around
the four cells bounding each edge of the lattice, 
the field was assumed to change values
by moving in the shorter of the two possible directions in the complex plane.
A cosmic string was then placed along an edge if the field traced out
a 360$^o$ circle in the complex plane as a path was traced around
the four cells bounding that edge.
We define an infinite string as one
which crosses all the way from one
end of the box to the other in at least one of the three directions.
Note that this definition
differs from
that used by Robinson and Yates \cite{RY}, who used a cutoff
in the string length.

Consider first the simplest case, for which we have no smoothing, and
the mean field value is set to zero.  This corresponds exactly to the 
simulations
of reference \cite{RY}.  To find the value of $f_\infty$ and its
variance, we performed 32 simulations
for each value of $n$, grouped into eight groups of four simulations.
Within each group of four, we derived an average value for $f_\infty$
by dividing the combined length of infinite strings in all four
simulations by the total length of string in that group.
This procedure corresponds to considering each of the
four simulations in a group as sampling a different region of space.
We then averaged $f_\infty$ for the eight groups of simulations to derive
a final mean $f_\infty$ and variance.  [Note that in deriving Figures 3-6,
we used 4 groups of 4 simulations rather than 8].

In Figure 1a, we present our results for no smoothing with zero mean field.
(All error bars are 1-$\sigma$).
These results agree closely with those of reference \cite{RY}, as they should.
This indicates that $f_\infty$ is relatively insensitive to the exact
definition of infinite string, since this is the only difference between
our simulations and those of reference \cite{RY}.

Now we repeat the simulations, but add a
random uniform background field $\phi_b$ 
to the entire box.  This background field has a Gaussian distribution
with a variance which can be determined, in principle, from equation (\ref{power}).
In practice, we use a simpler method:  we embed our
simulation volume in a larger $128^3$ box
with the same power spectrum, in which our entire
simulation volume occupies a single cell of the larger box.
The mean field $\phi_b$ in the simulation volume
is then simply the value of $\phi$ in a single
(randomly-chosen) cell of the larger box.  Note that this method
still causes a loss of power at the largest scales (i.e., Fourier modes
with wavelengths longer than the size of the larger box), but it provides
an effective dynamic range of $128^2 = 16,384$, which represents a
considerable improvement over the zero mean field case.

In Figure 1b, we show our results for
this case.  The greatest difference between Figures 1a and 1b
occurs for $n \le -2$:
the addition of the background field
produces a larger variance between runs, and it
reduces $f_\infty$ sharply.
The graph in Figure 1b is
consistent with the suggestion
of reference \cite{RY} that $f_\infty$ goes to zero when $n$
is less than some critical value $n_c$, where
$n_c>-3$; i.e., the average string density is still non-zero
when the infinite strings disappear.  In reference \cite{RY},
it was claimed that $n_c = -2$, but this is inconsistent
with our results.  If we exclude points for which $f_\infty$
is more
than 3-$\sigma$ from zero, then we find $n_c < -2.2$.
Although our results are consistent with the possibility that
$n_c > -3$, they do not rule out the possibility that
$n_c = -3$ (i.e., the infinite strings do not disappear until
all the strings disappear).

In Figure 2, we plot the
total length in closed loops and infinite strings as a function
of $n$ with the inclusion of the background field.
(All of the results presented in this paper, with the exception of Figure 1a, include
this background field).
Although
the length in infinite strings is significantly reduced below
that obtained without a background field (e.g., see \cite{RY}),
we cannot tell with certainty at which value of $n$ the infinite
strings disappear.

In Figure 3, we show the size distribution of closed loops
for two representative cases ($n = 0$ and $n=-2.8$) when
the background field is included.  In both cases, $N(l)$ follows
a power law in $l$, but the slope changes noticeably with $n$.
Fitting the points with $10 \le l \le 100$, we find that
\begin{eqnarray}
N(l) &\propto& l^{-2.62 \pm 0.01},~~~~(n=0),\\
N(l) &\propto& l^{-2.35 \pm 0.04},~~~~(n=-2.8).
\end{eqnarray}
These fits are shown as dashed lines in Figure 3.
Neither of these is consistent with the frequently assumed \cite{VV,SF,RY}
behavior $N(l) \propto l^{-2.5}$, but even the original
($n=0$) simulations of Vachaspati and Vilenkin \cite{VV} found that
$N(l) \propto l^{-2.6 \pm 0.1}$.

Now consider the effect of smoothing the $\phi$ field on small
scales.  For a window function $W(\rvec)$, the smoothed field
$\phi_s(\xvec)$ is the convolution of $\phi(\xvec)$ with $W(\rvec)$:
\begin{equation}
\phi_s(\xvec) = \int d^3 r ~\phi(\xvec+\rvec) W(\rvec).
\end{equation}
The effect of smoothing is to reduce the magnitude of the small-scale
fluctuations by averaging them out over the window function.
We calculate $f_\infty$ when $\phi$ is smoothed
with three different window
functions, the spherical Gaussian window function,
\begin{equation}
\label{gauss}
W(r) = \exp(-r^2 / 2 r_0^2),
\end{equation}
the spherical tophat,
\begin{eqnarray}
\label{tophat}
W(r) &=& 1~~~(r < r_0), \nonumber \\
W(r) &=& 0~~~(r > r_0),
\end{eqnarray}
and a sharp cut-off in $k$-space, which corresponds,
in physical space, to the smoothing
\begin{equation}
W(r) = {\sin(r/r_0) \over (r/r_0)^3} - {\cos(r/r_0) \over (r/r_0)^2},
\end{equation}
where we then normalize each window function to give $\int W(r) d^3 r = 1$.

We now examine
the variation of
$f_\infty$ with $r_0$.  We consider only
the case $n=0$ because it resembles the most likely
scenario for string formation (i.e., no long-range correlations in the $\phi$
field), and for the range of $n$ values we have examined, it has the most
short-range power and should therefore show the largest sensitivity to
smoothing.  Our results are shown in Figure 4a for the spherical Gaussian,
in Figure 4b for the spherical tophat, and in Figure 4c for the sharp
cut-off in $k$-space.  

Note that $r_0$ represents something different
for each of our three window functions, so it is meaningless to
compare $f_\infty$ for the same value of $r_0$ in each figure.
However, the important point is whether there exists a range of
values of $r_0$ for a given window function which produces a relatively
constant set of values for $f_\infty$.
For the case of Gaussian smoothing, $f_\infty$ decreases with $r_0$
but eventually ``plateaus" for  $r_0 \ge 1.5$,
suggesting that the value of $f_\infty$ in this
region is the ``true" value of $f_\infty$ for Gaussian smoothing.
Averaging $f_\infty$ for
$1.5 \le r_0 \le 2.5$, we obtain $f_\infty = 0.71 \pm 0.01$.
As $r_0$ is increased, the size of the smoothing
volume grows compared to the size of the box, and we see larger
fluctuations from one run to the next.  However, the rms fluctuations
in $f_\infty$
remain within $10\%$ of our mean value ($f_\infty = 0.71$)
over the entire range of $r_0$ values we have examined ($r_0 \le 10)$.
An
average of $f_{\infty}$ over the range $1.5\le r_0\le 10$ gives
$f_{\infty}=0.72 \pm 0.04$.

For spherical tophat smoothing,
we again see that $f_\infty$ decreases with $r_0$, but no obvious plateau is
visible in Figure 2b.  In fact, tophat smoothing fails to eliminate
lattice effects.  If we substitute the smoothed version
of the power spectrum
into equation (\ref{k2}), we find that for
the spherical tophat window function with
$n=0$, the integral in the numerator in equation
(\ref{k2}) fails to converge at $k \rightarrow \infty$, indicating
that it is the cubic lattice, rather than the spherical tophat, which
provides the cutoff in this case.  This is confirmed by the loop
distribution for the spherical cutoff; unlike our other two window
functions, the spherical tophat produces
a loop distribution which retains a large number of loops
with size of order the cell size.

The sharp $k$-space cut-off shows qualitatively similar behavior to the
Gaussian window function.  The value of $f_\infty$ decreases slightly 
as a function of $r_0$, and reaches a plateau for $r_0 \ge 1$. 
Averaging $f_\infty$ for $1 \le r_0 \le 3$ and $1\le r_0\le 10$, we get
$f_\infty = 0.82 \pm 0.01$ and $f_{\infty}=0.85\pm 0.05$, respectively.
 
In Figure 5, we demonstrate
the effect of smoothing on the size distribution of the loops,
by showing $N(l)$ for the closed loops with Gaussian smoothing and $r_0 = 3$.
Note that smoothing does not eliminate the closed loops smaller than the
size of the window function; rather, it reduces the number of loops to
a constant as a function of $l$ for small $l$.  Of course, the total length in
closed loops is then dominated by loops with size near the point
at which the power law behavior begins (in this case, $l \approx 30$).

Smoothing allows us to test the analytic formula for the total
string density given in equations (\ref{L/V}) and (\ref{k2}).
For the power spectrum given by equation (\ref{power}) smoothed with
the Gaussian window function given by equation (\ref{gauss}),
the expression for $L/V$ reduces to
\begin{equation}
\label{analytic}
L/V = {1 \over \pi r_0^2} \Biggl({n+3 \over 3} \Biggr).
\end{equation}
(Note that in this case, it is obvious that ${L/V} \rightarrow 0$ as
$n \rightarrow -3$).
To compare with our simulations on the lattice, $L/V$ must be multiplied
by the number of cells (128$^3$ in our case) times the number of
independent lattice edges per cell (3 for a cubic lattice).  The
expected string length derived in this way from equation (\ref{analytic})
is compared to our simulation results in Figure 6, for a smoothing
length $r_0 = 3$.  The agreement is excellent, within the expected
statistical fluctuations.  Whether one views this as a confirmation
of the analytic expression or of the validity of our numerical
simulations is a matter of personal preference.

\section{Discussion}

Our results are consistent with the possibility that
a transition occurs in the string network at some value $n= n_c$,
with $-3 \le n_c < -2.2$; when $n < n_c$ the infinite
strings disappear from the network, leaving only closed loops.
Robinson and Yates
\cite{RY} argued that $n_c=-2$, but
this value is not consistent with our results.
Furthermore, we cannot rule out the possibility that
$n_c = -3$, i.e., the infinite strings do not disappear until
the mean string density goes to zero.

In fact, the following argument suggests that infinite
strings should not disappear for $n > -3$.
The real and imaginary parts of $\phi$ are independent Gaussian fields,
and the zeros of each of these fields form a set of two-dimensional
surfaces.  Consider first the surfaces defined by Re$(\phi) = 0$.
Since the volumes of space occupied by the regions with positive
and negative Re($\phi$) are equal, we expect both regions
to percolate to an infinite distance. Hence, the boundaries
dividing these regions should contain at least some infinite
surfaces.
This argument holds for both the surfaces
defined by Re($\phi) = 0$ and Im($\phi) = 0$.  The intersection
of these two sets of surfaces gives us the location of the cosmic strings.
Since infinite surfaces must exist in both sets of surfaces, it
appears that there
must also be infinite cosmic strings.  Note that it is possible
to imagine rather arcane distributions of the fields which violate
this argument.  For example, the regions with Re$(\phi) > 0$ and
Re$(\phi) < 0$ could be nested inside of each other in larger and
large finite volumes, producing a fractal distribution with
arbitrarily large but finite surfaces of Re$(\phi) = 0$.
However, it seems unlikely that a Gaussian field could lead to
such a distribution.  The crucial point in this argument
is the fact that the distribution is symmetric with respect
to positive and negative values of Re($\phi$) and Im($\phi$);
if we relax this assumption and ``bias" the distribution,
the argument no longer holds.  In fact, infinite strings are observed
to disappear in simulations with such a ``bias"
\cite{vachaspati}, \cite{kibble}, \cite{HS}.

The aim of our simulations with a smoothed field $\phi$ was to obtain
a lattice-independent estimate of $f_{\infty}$.  We considered only the
white noise spectrum, $n=0$, and found $f_{\infty}\approx 0.7$ for
spherical Gaussian smoothing and $f_{\infty}\approx 0.8$ for a sharp
cut-off in $k$-space.  These values are comparable to those obtained
in earlier lattice simulations without smoothing \cite{VV,SF}.  The
variation of $f_{\infty}$ for different choices of smoothing is not
surprising, since, for example, the total string density $L/V$ in
equations (\ref{L/V}) and (\ref{k2}) clearly depends on the
short-wavelength behavior of the spectrum $P(k)$ (which is affected by
the smoothing).  This variation is again comparable to the variation
between the values of $f_{\infty}$ in simulations without smoothing on
different types of lattice \cite{VV}, \cite{SF}, \cite{HS}.

We find no evidence
supporting the hypothesis \cite{Borrill} that the presence of infinite
strings is due entirely to the lattice and that they should
disappear in lattice-free simulations.  The smoothing length we used
was sufficiently large to ensure that the loops making the largest
contribution to the total string length had sizes much greater than
the lattice cutoff, so that the lattice effects were minimal.  Still,
we found a substantial fraction of the total length to be in infinite
strings. 

We are currently investigating the formation of domain walls and monopoles
with correlated fields.
Given that a parallel literature on this subject
exists in condensed matter physics \cite{bradley},
these results may have applications beyond the purely cosmological.
It is conceivable, for example, that the decline of the infinite
string density with
the decrease of the spectral index $n$ can be tested experimentally.
What one needs is a condensed matter system (such as liquid He$^4$) in
which linear defects are formed at a second-order phase transition.
Defect formation can then be observed by a rapid temperature (or
pressure) quench from above to below the transition point
\cite{Zurek,Hendry}. 
Near the critical temperature $T_c$, the order parameter develops 
long-range fluctuations.  At $T=T_c$, the
fluctuation spectrum is a power law of the form (\ref{power}) with
$n=-2+\eta$, where the critical exponent $\eta$ is typically a small
number, $\eta\lesssim 0.05$ \cite{Huang} ($\eta\approx 0.05$ for He$^4$).  
If the system is allowed to equilibrate
very close to the critical point and is then rapidly quenched to
subcritical temperatures, one can expect the length in infinite
strings to be suppressed  compared to a quench from a
temperature well above $T_c$ (where the fluctuation spectrum is close
to $n = 0$).  Our Fig. 2 suggests a suppression roughly by an order of
magnitude, while the value of $n_c=-2$ conjectured by Robertson and
Yates \cite{RY} would give a much more dramatic suppression.  It
should be noted that a realistic quench is a rather complicated
process, and its outcome can depend on a variety of physical effects
(for a recent discussion see \cite{Zurek,K}).

\acknowledgements

R.J.S. is supported by the Department of Energy
(DE-AC02-76-ER01545) and by NASA
(NAG 5-2864 and NAG 5-3111).  A.V. is supported by the National
Science Foundation.

%
%

\vfill
\eject

\centerline{\bf FIGURE CAPTIONS}

\vskip 0.5 cm

\noindent Figure 1:  The fraction of total string length
in the form of infinite strings, $f_\infty$, as a function of $n$, where
$P(k) \propto k^n$, and $P(k)$ is the power spectrum of the complex
Gaussian
field which gives rise to the strings, where (a)  the field
in the box is forced to have zero mean, and (b) a background
non-zero mean
field is added to the box to simulate the effects of long-wavelength
modes.

\vskip 0.5 cm

\noindent Figure 2:  The total length $L$ in closed loops (open squares)
and infinite strings (crosses) as a function of $n$, where
$P(k) \propto k^n$, and $P(k)$ is the power spectrum of the complex
Gaussian
field which gives rise to the strings.  A background non-zero
mean field is added to the box to simulate the effect
of long-wavelength modes.

\vskip 0.5 cm

\noindent Figure 3:  The number of loops $N$ with a given
length $l$ for (a) $n = 0$ and (b) $n=-2.8$, where
$P(k) \propto k^n$, and $P(k)$ is the power spectrum of the complex
Gaussian
field which gives rise to the strings.  A background non-zero
mean field is added to the box to simulate the effect
of long-wavelength modes.  The dashed line is the best-fit power law
in each case.

\vskip 0.5 cm
 
\noindent Figure 4:  The fraction of total string length
in the form of infinite strings, $f_\infty$, as a function
of smoothing length $r_0$ for (a) Gaussian smoothing,
(b) spherical tophat smoothing, (c) a sharp $k$-space cutoff,
where the power law index is $n=0$ (no field correlations), and
a background non-zero
mean field is added to the box to simulate the effect
of long-wavelength modes.

\vskip 0.5 cm

\noindent Figure 5:  The number of loops $N$ with a given
length $l$ for a field with $n=0$ (no field correlations)
smoothed with a Gaussian window function with smoothing length
$r_0 = 3$.  A background non-zero mean field is added to the box to simulate
the effect of long-wavelength modes.

\vskip 0.5 cm

\noindent Figure 6:  The total string length $L$ as a function of $n$, where
$P(k) \propto k^n$, and $P(k)$ is the power spectrum of the complex
Gaussian
field which gives rise to the strings.
The points with error bars are the results of our simulation, and
the solid line is the analytic prediction.
A background non-zero mean field is added to the box to simulate
the effect of long-wavelength modes.

\vfill
\eject

\centerline{\epsfbox{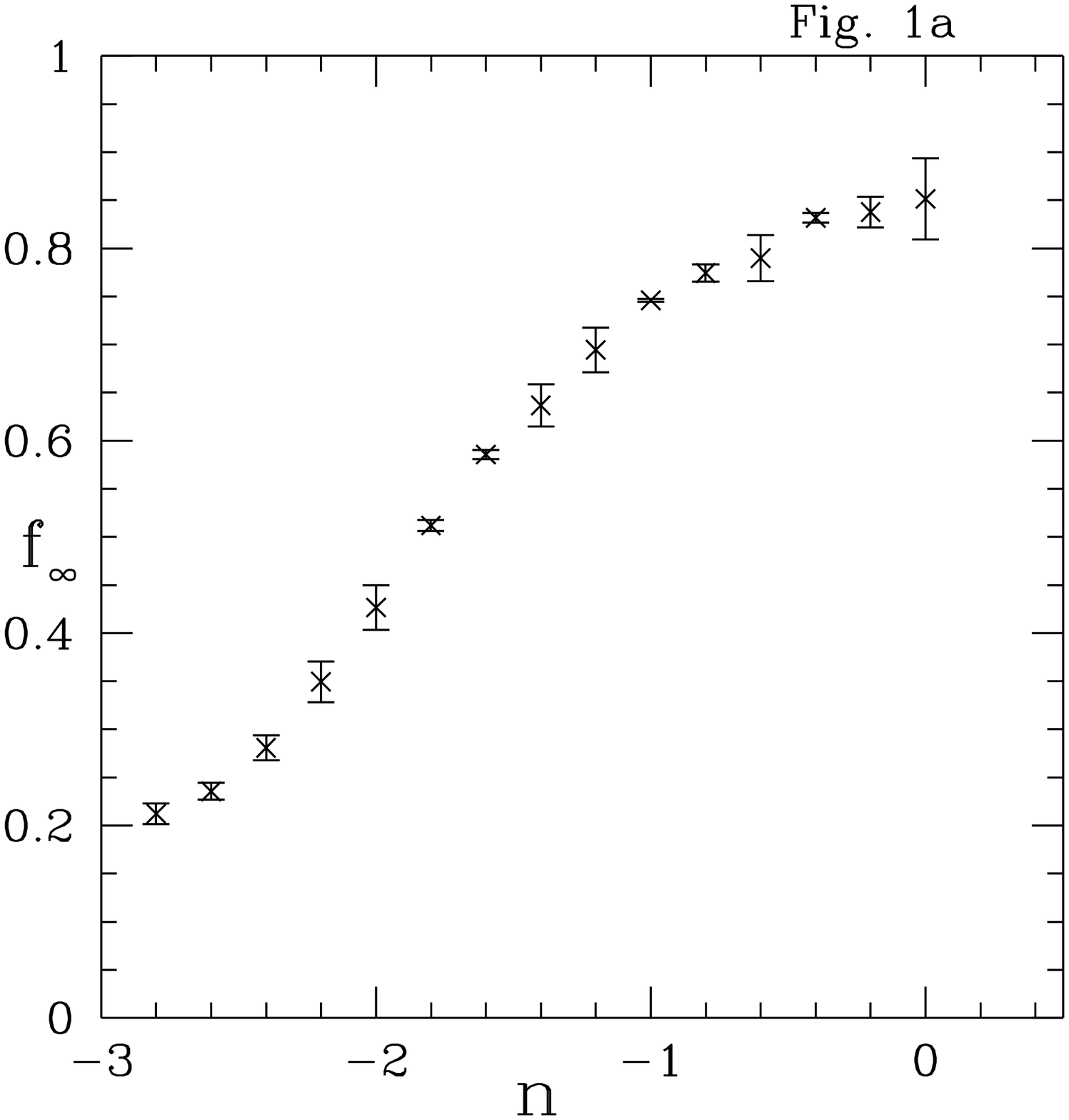}}

\centerline{\epsfbox{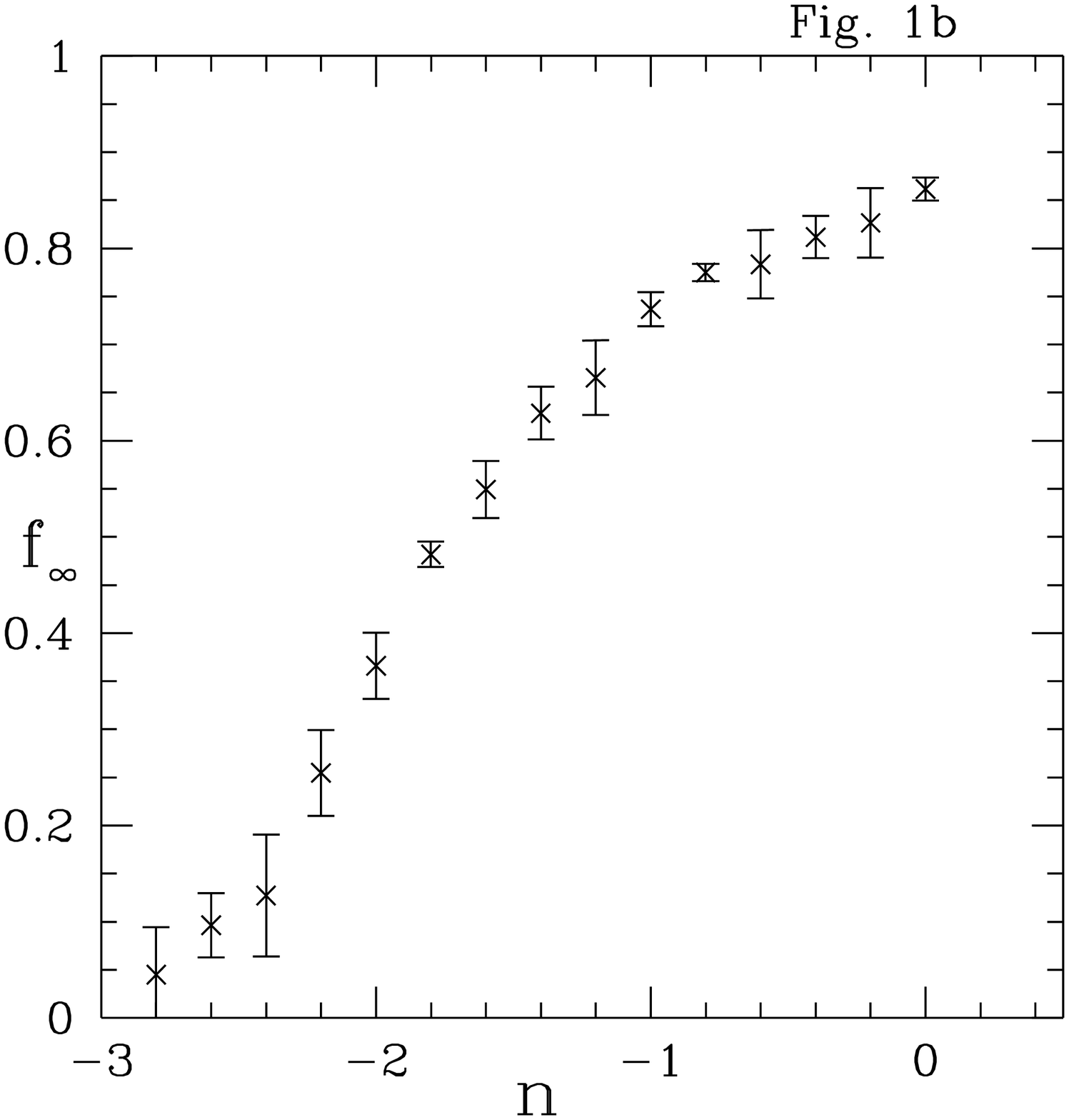}}

\centerline{\epsfbox{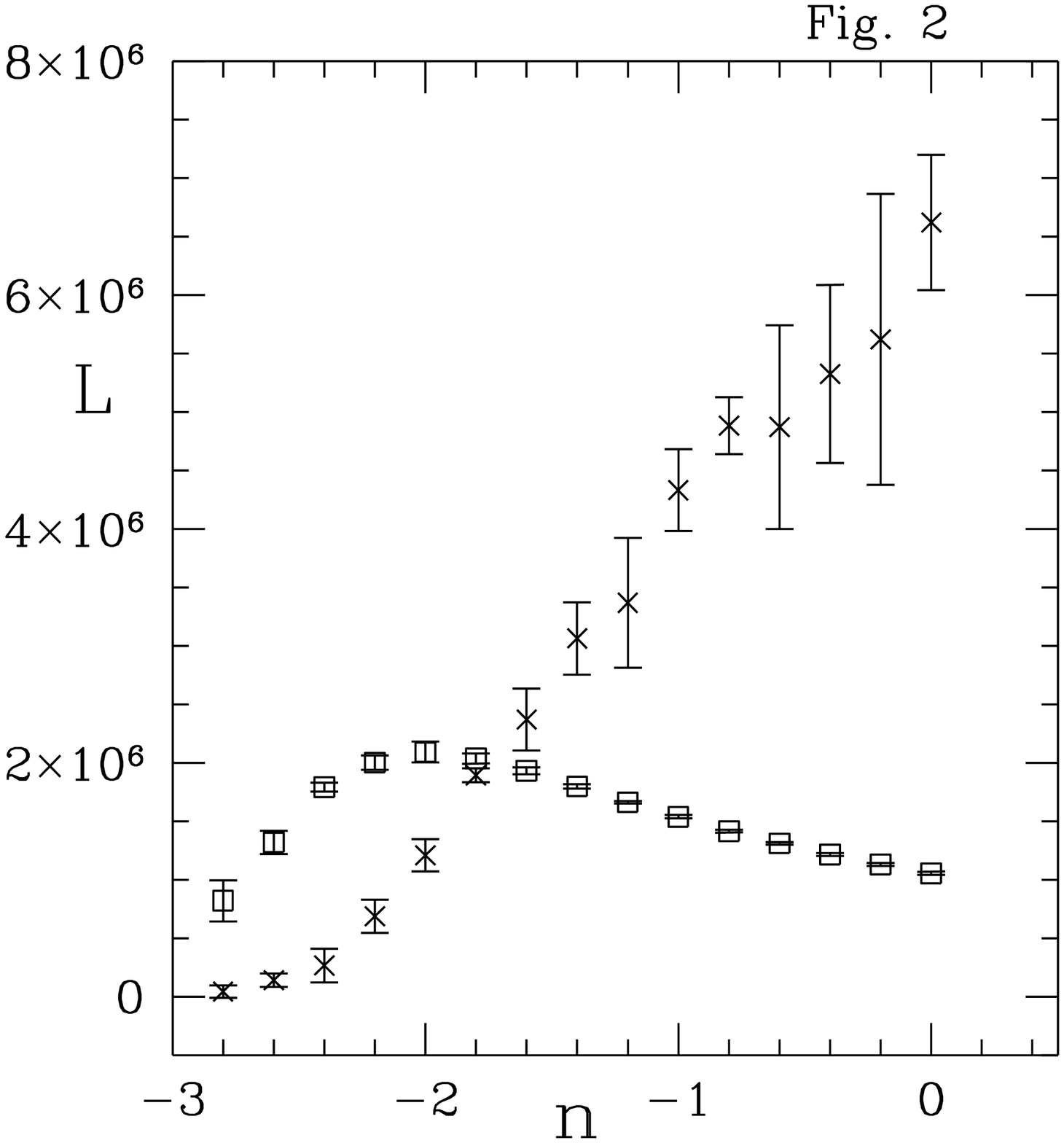}}

\centerline{\epsfbox{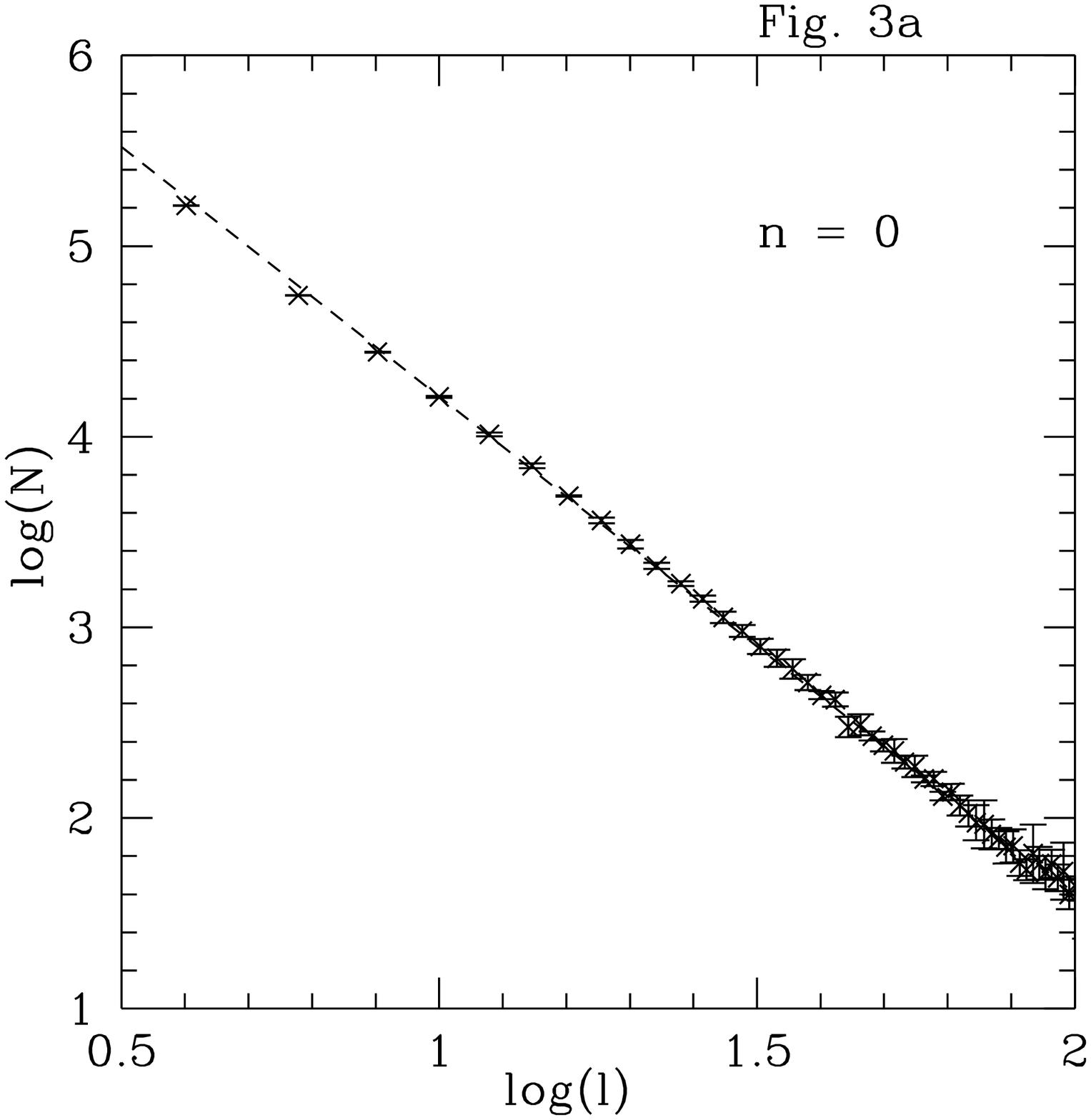}}

\centerline{\epsfbox{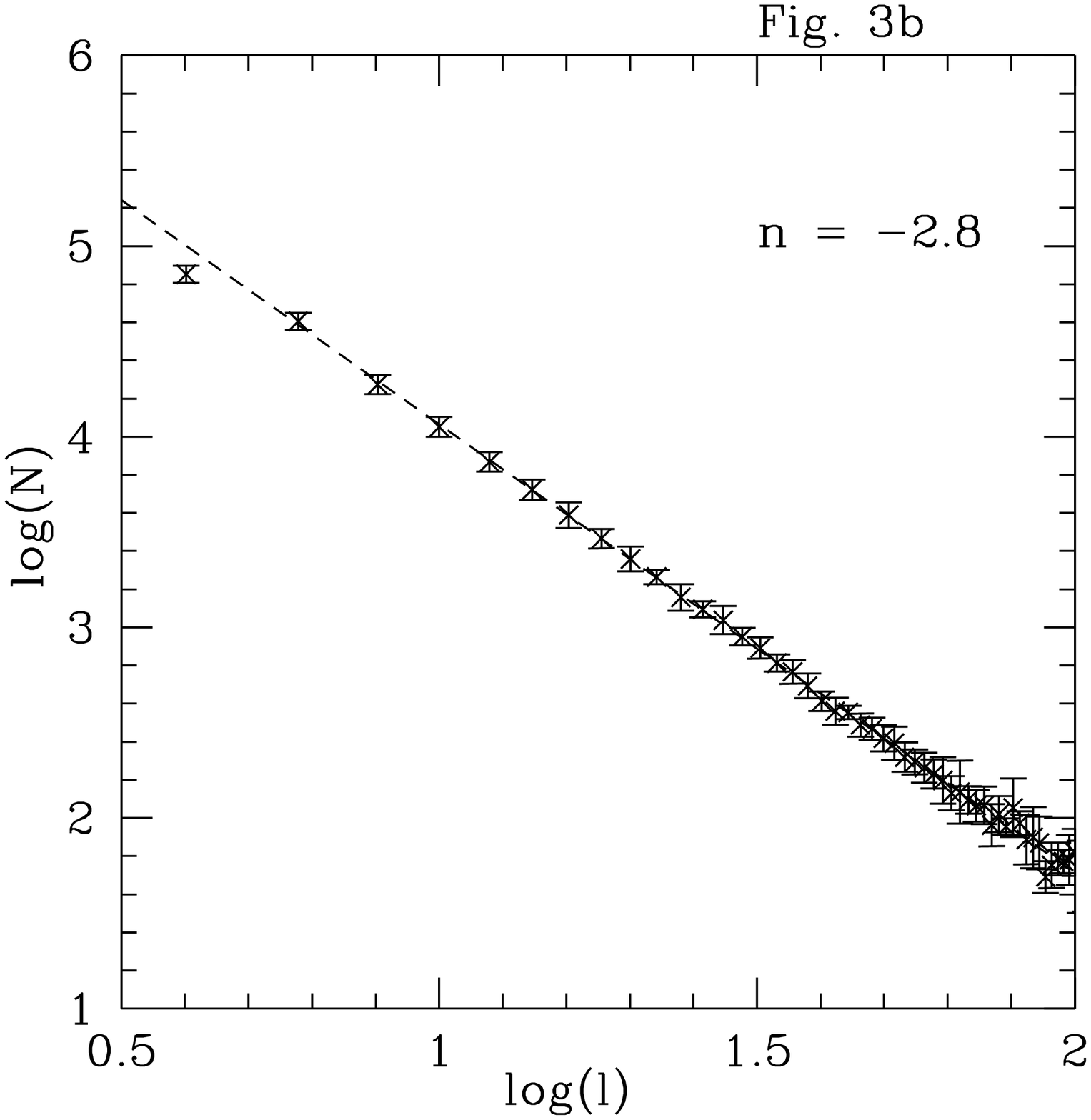}}

\centerline{\epsfbox{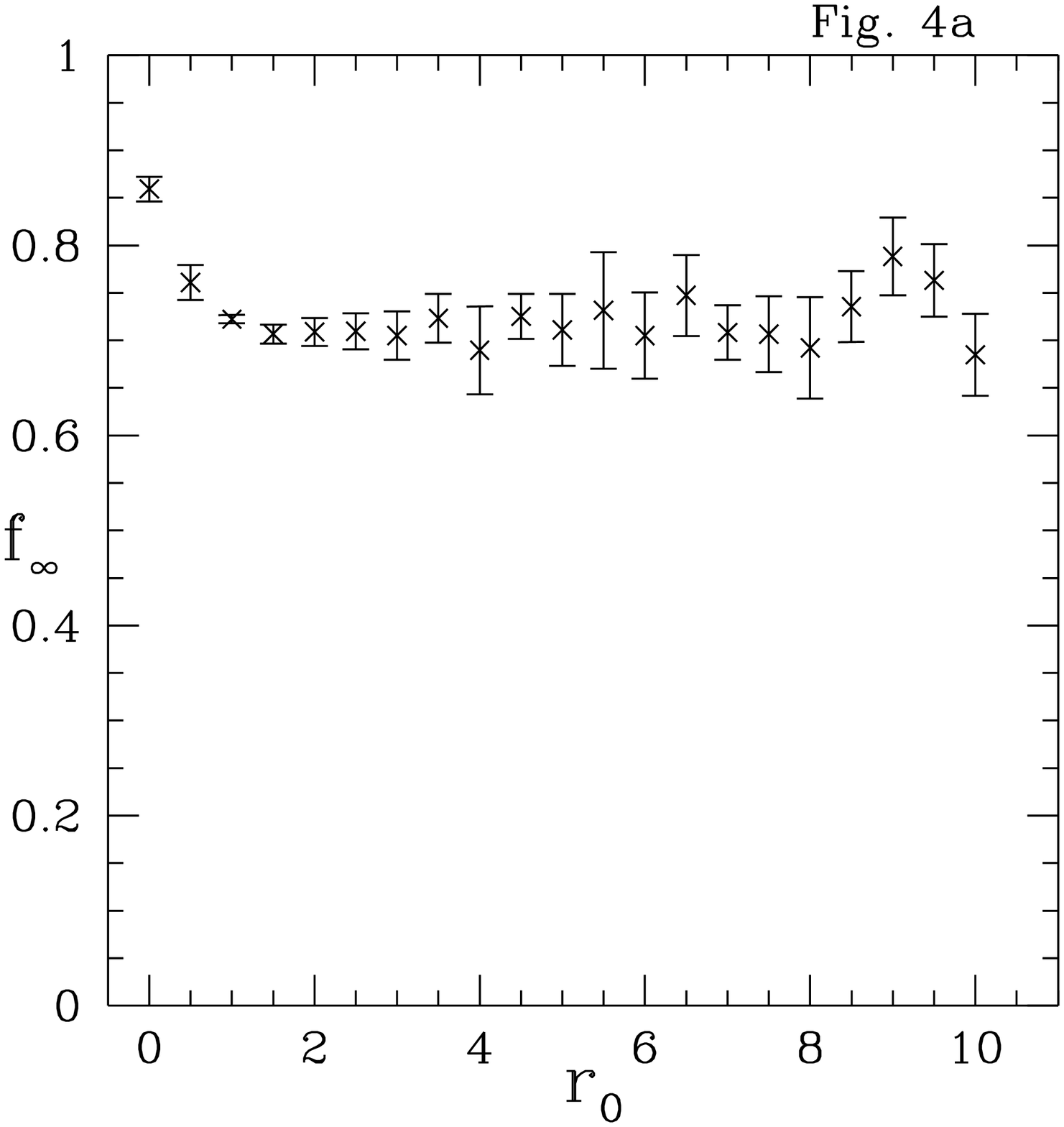}}

\centerline{\epsfbox{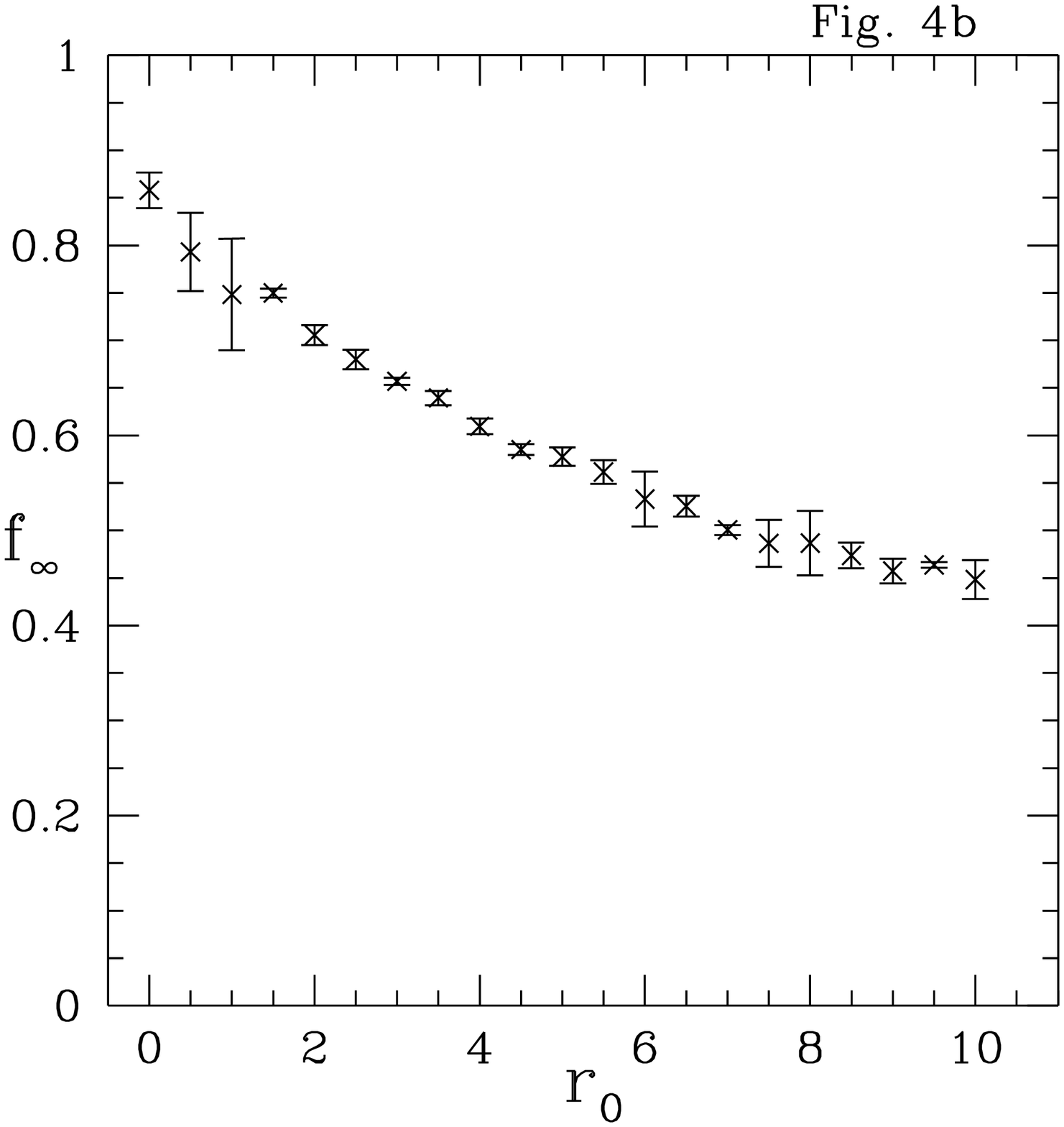}}

\centerline{\epsfbox{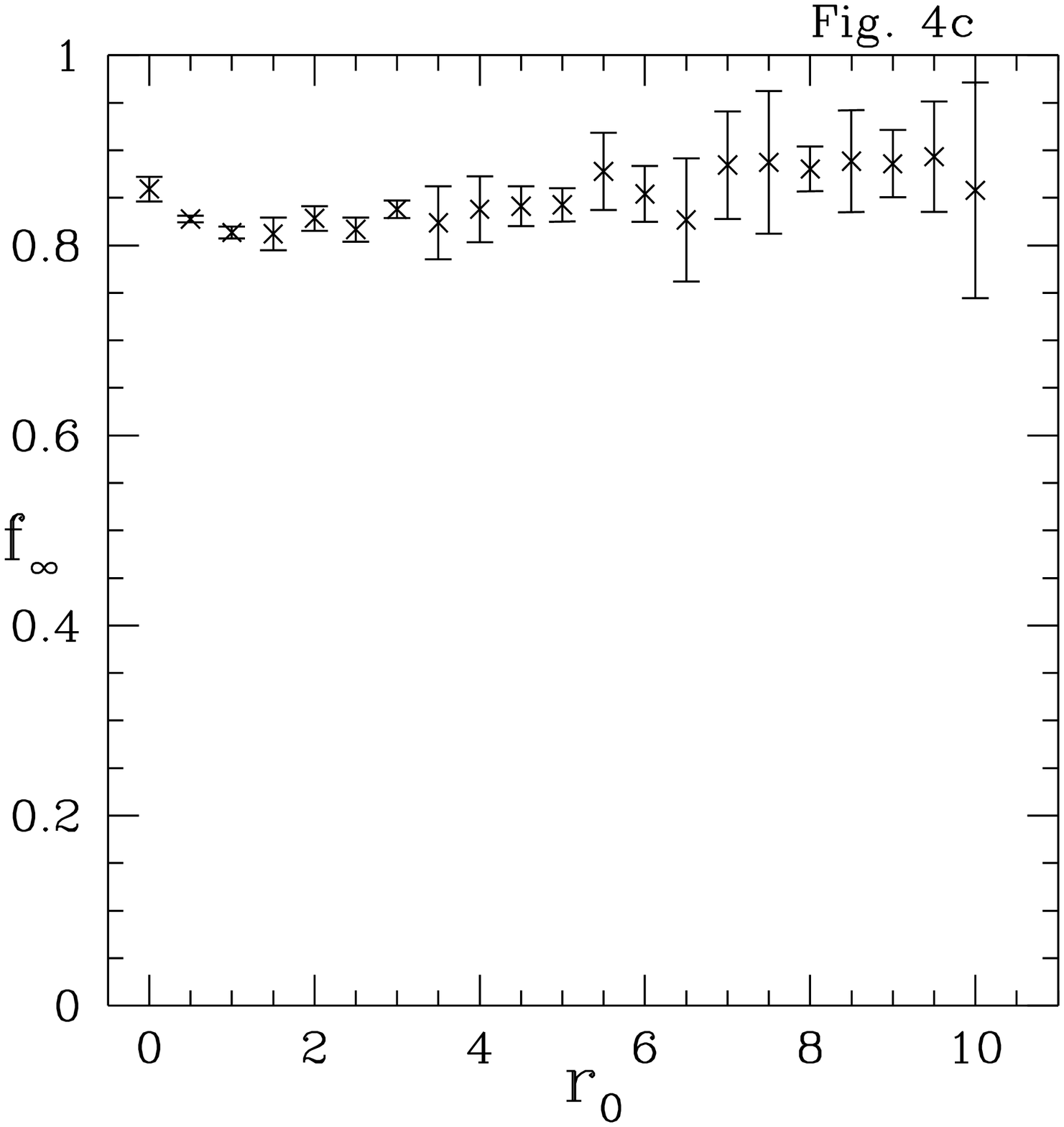}}

\centerline{\epsfbox{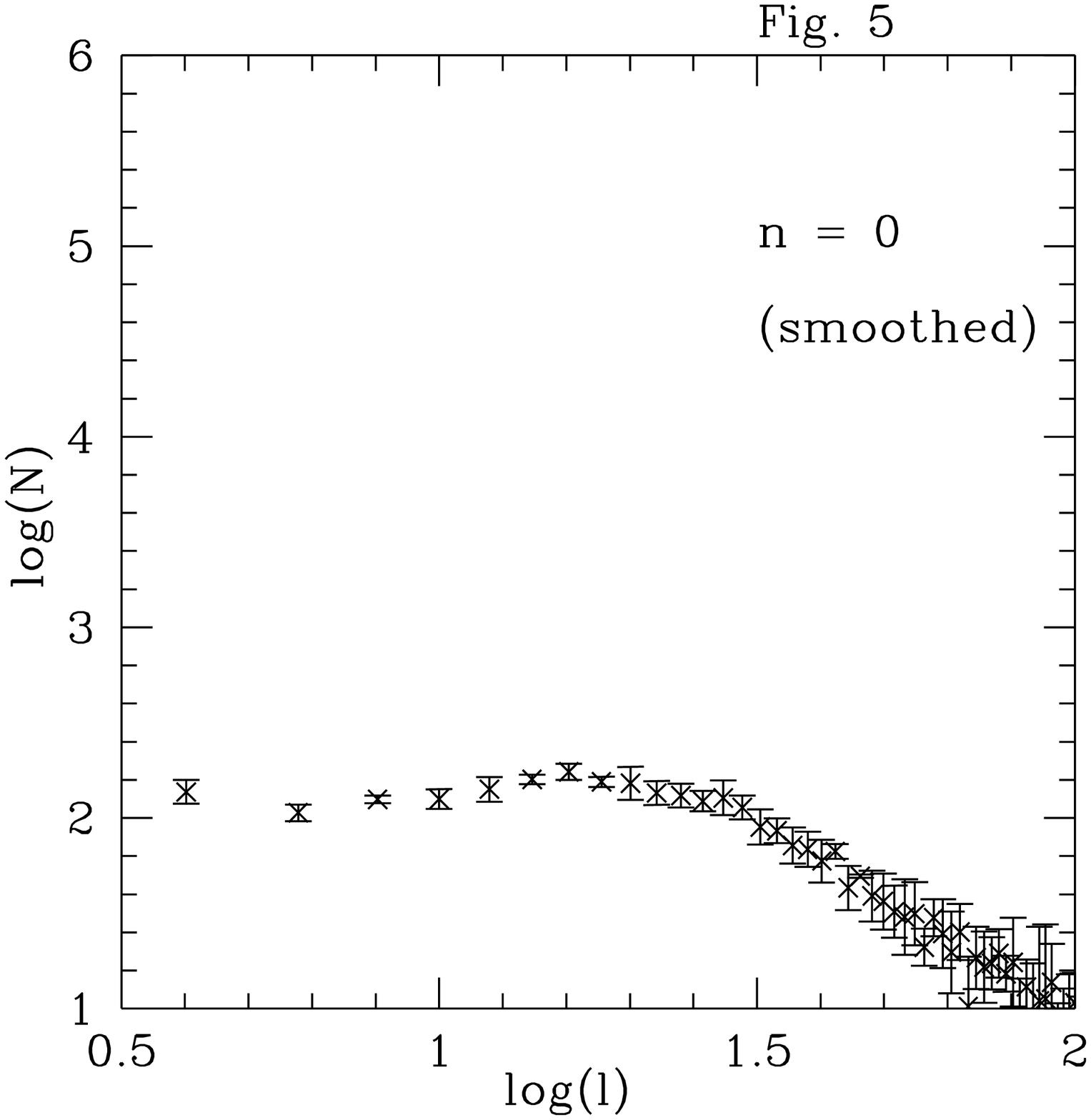}}

\centerline{\epsfbox{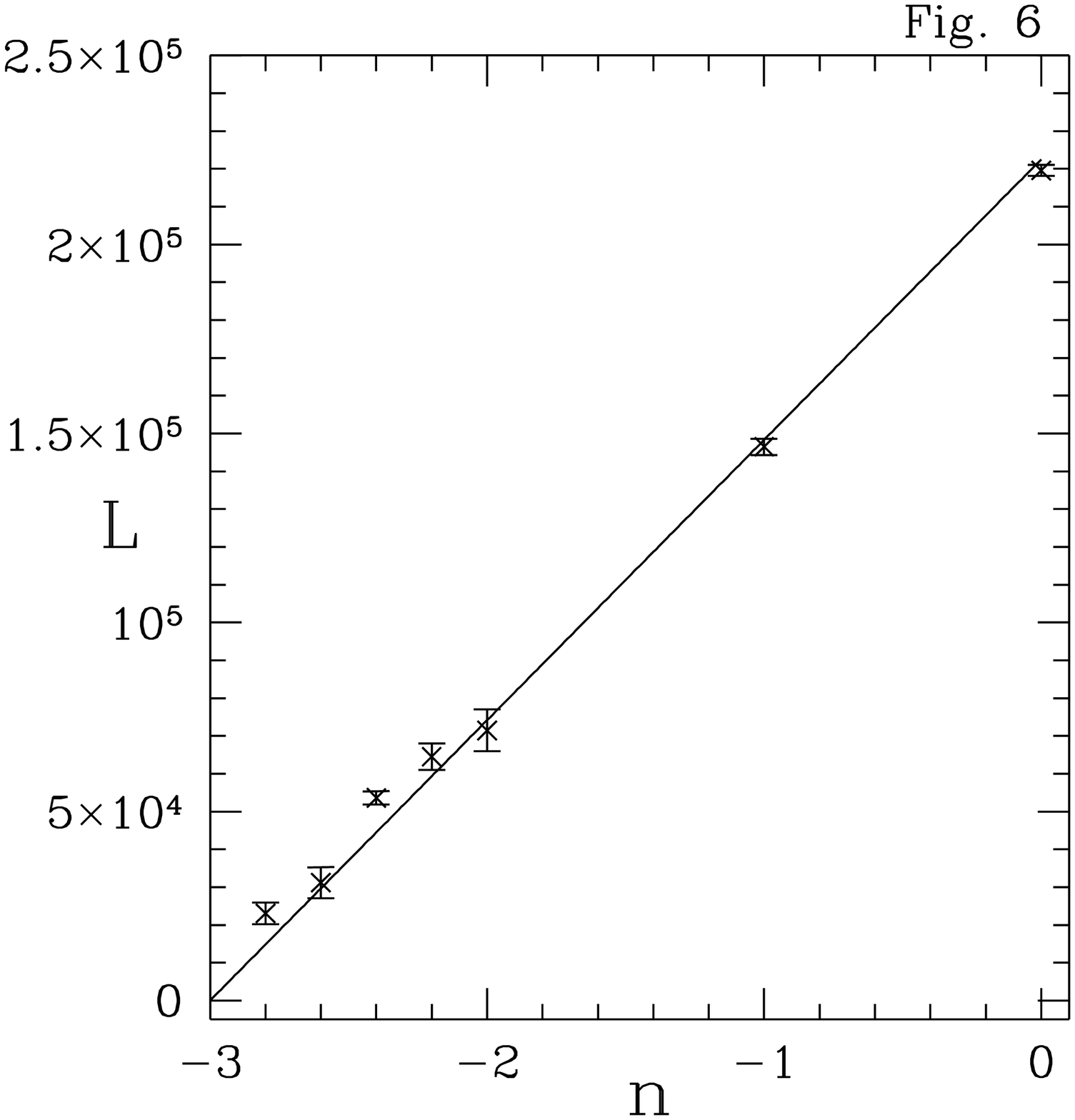}}


\begin{thebibliography}{99}

\bibitem{review}
A. Vilenkin and E.P.S. Shellard, {\it Cosmic Strings and
other Topological Defects}, Cambridge University Press, Cambridge,
1994;  M.B. Hindmarsh and T.W.B. Kibble, Rep. Prog. Phys. {\bf 55}, 478
(1995)

\bibitem{condmat}
See, e.g., I. Chuang, R. Durrer, N. Turok and B. Yurke, Science {\bf
251}, 1336 (1991); G.E. Volovik and T. Vachaspati, cond-mat/9510065,
and references therein.

\bibitem{VV}
T. Vachaspati and A. Vilenkin,
Phys. Rev. D {\bf 30}, 2036 (1984).

\bibitem{SF}
R. J. Scherrer and J. Frieman,
Phys. Rev. D {\bf 33}, 3556 (1986).

\bibitem{vachaspati}
T. Vachaspati,
Phys. Rev. D {\bf 44}, 3723 (1991).

\bibitem{kibble}
A. Yates and T.W.B. Kibble,
Phys. Lett. B {\bf 364}, 149 (1995).

\bibitem{HS}
M. Hindmarsh and K. Strobl,
Nucl. Phys. B {\bf 437}, 471 (1995).

\bibitem{Borrill}
J. Borrill,
Phys. Rev. Lett. {\bf 76}, 3255 (1996).

\bibitem{RY}
J. Robinson and A. Yates,
Phys. Rev. D {\bf 54}, 5211 (1996).

\bibitem{VOS}
E. T. Vishniac, K. A. Olive, and D. Seckel,
Nucl. Phys. B {\bf 289}, 717 (1987).

\bibitem{bradley}
R. M. Bradley, J.-M. Debierre, and P.N. Strenski,
J. Phys. A {\bf 25}, L541 (1992);
R.M. Bradley, P.N. Strenski, and J.-M. Debierre,
Phys. Rev. A {\bf 45}, 8513 (1992).

\bibitem{Zurek}
W. H. Zurek, Cosmological experiments in condensed matter systems,
cond-mat/9607135, Phys. Rep., in press.

\bibitem{Hendry}
P. C. Hendry {\it et. al.}, Nature {\bf 368}, 315 (1994).

\bibitem{Huang}
See, e.g., K. Huang, {\it Statistical Mechanics}, Wiley, New York,
1987.

\bibitem{K}

A.J. Gill and T.W.B. Kibble, cond-mat/9603050, unpublished.

\end{thebibliography}
\end{document}